\documentclass[aps,prl,twocolumn,superscriptaddress]{revtex4-1}

\usepackage[english]{babel}
\usepackage[latin1,utf8]{inputenc}
\usepackage{amsmath}
\usepackage{amsthm}
\usepackage{amssymb}
\usepackage{graphicx}

\newcommand{\Fig}[1]{Fig.~\ref{fig:#1}}
\newcommand{\Eq}[1]{Eq.~\eqref{eq:#1}}
\newcommand{\eq}[1]{\eqref{eq:#1}}
\newcommand{\Eqs}[2]{Eqs.~\eqref{eq:#1} \& \eqref{eq:#2}}
\newcommand{\Eql}[2]{Eqs.~\eqref{eq:#1} -- \eqref{eq:#2}}

\newcommand{\vect}[1]{{{\vec{#1}}}}
\newcommand{\EV}{\epsilon} 	
\newcommand{\EVtmp}{k}		

\usepackage{color}
\definecolor{myOrange}{rgb}{1,0.5,0.1}
\definecolor{myRed}{rgb}{0.8,0.1,0.1}
\definecolor{myGreen}{rgb}{0.7,0,0.8}
\definecolor{myGray}{rgb}{0.6,0.6,0.6}

\definecolor{light-gray}{gray}{0.95}

\usepackage[normalem]{ulem}

\begin{document}

\title{Mimicking inflation with 2-fluid systems in a strong gradient magnetic field}
\author{Zack Fifer}
\affiliation{School of Mathematical Sciences, University of Nottingham, University Park, Nottingham NG7 2RD, UK}
\affiliation{Centre for the Mathematics and Theoretical Physics of Quantum Non-Equilibrium Systems, University of Nottingham, Nottingham NG7 2RD, UK}
	
\author{Theo Torres}
\affiliation{School of Mathematical Sciences, University of Nottingham, University Park, Nottingham NG7 2RD, UK}
\affiliation{Centre for the Mathematics and Theoretical Physics of Quantum Non-Equilibrium Systems, University of Nottingham, Nottingham NG7 2RD, UK}
	
\author{Sebastian Erne}
\affiliation{School of Mathematical Sciences, University of Nottingham, University Park, Nottingham NG7 2RD, UK}
\affiliation{Centre for the Mathematics and Theoretical Physics of Quantum Non-Equilibrium Systems, University of Nottingham, Nottingham NG7 2RD, UK}
	
\author{Anastasios Avgoustidis}
\affiliation{School of Physics and Astronomy, University of Nottingham, Nottingham NG7 2RD, UK}
	
\author{Richard J. A. Hill}
\affiliation{School of Physics and Astronomy, University of Nottingham, Nottingham NG7 2RD, UK}
	
\author{Silke Weinfurtner}
\affiliation{School of Mathematical Sciences, University of Nottingham, University Park, Nottingham NG7 2RD, UK}
\affiliation{Centre for the Mathematics and Theoretical Physics of Quantum Non-Equilibrium Systems, University of Nottingham, Nottingham NG7 2RD, UK}

	\begin{abstract}
		In the standard cosmological picture the Universe underwent a brief period of near-exponential expansion, known as Inflation. This provides an explanation for structure formation through the amplification of perturbations by the rapid expansion of the fabric of space. Although this mechanism is theoretically well understood, it cannot be directly observed in nature.
		We propose a novel experiment combining fluid dynamics and strong magnetic field physics to simulate cosmological inflation. 
		Our proposed system consists of two immiscible, weakly magnetised fluids moving through a strong magnetic field in the bore of a superconducting magnet. By precisely controlling the propagation speed of the interface waves, we can capture the essential dynamics of inflationary fluctuations: interface perturbations experience a shrinking effective horizon and are shown to transition from oscillatory to squeezed and frozen regimes at horizon crossing.  		 
	\end{abstract}
	\maketitle

The physics of the Early Universe is a fascinating subject. In the standard scenario the primordial universe expanded in a nearly exponential fashion, a phase known as inflation. Originally designed to tackle outstanding puzzles of the celebrated Big Bang model (namely the horizon, flatness, and monopole problems~\cite{Guth,LindeNewInf}), it was soon realised that inflation provides much more. It explains the very origin of cosmic structure in an unexpected, but truly elegant and exotic way~\cite{MukhChib,Sasaki,GuthPi,Hawking,Starobinsky,BardSteinTurn}.
To resolve the horizon and flatness problems, inflation proposes that each of the three spacial dimensions in our Universe expanded by a factor $\!\sim\! e^{60}$ within a time interval $\sim\!\! 10^{-33}\ {\rm s}$. During this rapid expansion fluctuations are stretched beyond the characteristic scale of the expansion (known as the Hubble scale), at which point they stop evolving in time (the modes are said to be frozen). By this process initially small perturbations get amplified and converted to density fluctuations, eventually leading to the observed large-scale structure of our universe. In this letter we propose a setup that allows us to simulate this characteristic process in a laboratory setup.

We consider a system of two immiscible liquids, a diamagnetic layer lying atop a paramagnetic one, that can be moved at a precisely controllable rate through a strong, spatially varying magnetic field generated by a superconducting solenoid or a Bitter magnet (c.f.\ Ref.~\cite{BAL15}). 
The body forces applied to the liquid sample amount to an effective gravitational force, whose time dependence can be controlled by the geometry of the magnetic field and the sample's velocity. 
Thus, the fluids are subject to a time-varying effective gravitational field. It is well-known~\cite{UNR81} that small perturbations on the surface of irrotational fluids experience an effective geometry whose properties are determined by the background flow. 
The corresponding metric in our system takes the form $g_{\mu\nu} \propto {\rm diag}(-c^2, 1,1)$, where $c$ is the propagation speed of the interface perturbations. 
A key property of surface waves is that the propagation speed depends on gravity, which in our system is rendered time-dependent. 
This provides the versatility to engineer analogue cosmological spacetimes and study in detail the dynamics of perturbations in various time-dependent backgrounds. 
We discuss our experimental setup, and show that within this system an inflationary regime arises naturally: the effective speed of propagation of surface waves can be tailored to decrease in an exponential fashion giving rise to a shrinking `effective horizon'. 
Thus, the interface fluctuations exit the horizon mode by mode, transitioning from uncorrelated oscillatory behavior to frozen and squeezed regimes, in direct analogy to inflationary perturbations. 

There has been growing interest recently in analogue gravity experiments, focusing mainly on effects arising from static~\cite{WEI11,belgiorno2010hawking,euve2016observation,steinhauer2016observation} and rotating~\cite{torres2017rotational,vocke2017rotating} black hole spacetimes. More relevant for our proposal are time-dependent spacetime geometries, which have been subject to  theoretical~\cite{fedichev2004cosmological,barcelo2003analogue,weinfurtner2009cosmological,jain2007analog,cha2017probing} and experimental~\cite{jaskula2012acoustic,eckel2017supersonically} studies. Our experiment aims at achieving the first experimental observation of mode freezing in an analogue system. This would demonstrate the universality and robustness of this effect beyond cosmology, over a broader range of disciplines within physics.

\begin{figure}[!t] 
  \includegraphics[width=0.9\columnwidth]{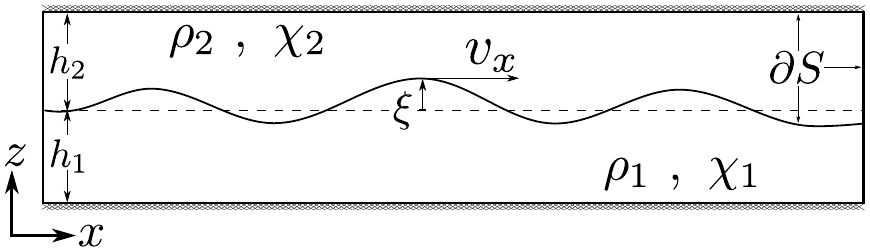}  
  \caption{Schematics of the two-fluid system. Two immiscible fluids, with densities $\rho_{1,2}$, magnetic susceptibility $\chi_{1,2}$, and height $h_{1,2}$, separated by the equilibrium interface at $z_0$ (dotted line). Gravity interface waves distort the interface layer (solid line) where each point is characterized by its amplitude $\xi$ and velocity $v_x$ ($v_y$). $\partial S$ are the boundaries (here depicted for fluid 2) given by a hard wall (upper), an interface boundary (right), and a moving boundary at the interface of the two fluids (lower).     } \label{fig:schematics}
\end{figure}

\textit{Two-fluid systems.}\textbf{---}We consider an immiscible two-fluid system (\Fig{schematics}), with densities $\rho_1 \!>\! \rho_2$, heights $h_{1,2}$, small magnetic susceptibilities $|\chi_{1,2}| \ll 1$, and flow velocities $v_{1,2}$. The system is subjected to a magnetic field $B(\vec{x},t) \sim 10\,$T with a large vertical field gradient $\sim 150\,$T/m. The flow of an inviscid and incompressible fluid is described by the continuity equation and Euler's momentum equation with the inclusion of the magnetic potential energy \cite{berry, geim, RichardArxiv}:
\begin{align} 
\vec{\nabla} \cdot \vec{v}_i &= 0  \label{eq:continuity} \\
\rho_i \left( \partial_{t} + \vec{v}_i \cdot\vec{\nabla} \right) \vect{v}_i &= \vect{\nabla} \left( -p_i + \frac{\chi_i}{2\mu_0}B^2 \right) + \rho_i \vec{g} \label{eq:euler} ~.
\end{align}
 The index $i=1$ ($i=2$) labels the lower (upper) fluid, $p_i$ is the fluid pressure, $\mu_0$ the vacuum  permeability, and $\vec{g} = (0,0,-g)$ is the acceleration due to gravity. The kinematic and, in case of a free surface, dynamic boundary conditions are 
\begin{align}
\vec{v}_i \cdot \vec{n} &= \vec{V} \cdot \vec{n} \qquad \text{on} \qquad \partial S \label{eq:BC_kinematic} \\
 [p] &= \sigma (R_1^{-1} + R_2^{-1}) \label{eq:BC_dynamic} ~,
\end{align}
on a boundary $\partial S$ with velocity $\vec{V}$. Equation~(\ref{eq:BC_kinematic}) states that the velocity of the fluid equals the velocity of the boundary along the outward normal $\vec{n}$ to the boundary. The angled bracket $\left[ * \right]$ denotes the jump in value across the interface, here the jump in pressure $p$ according to the Young-Laplace law \cite{landaufluid} with surface tension $\sigma$, and principal radii of curvature $R_{1,2}$. 

We assume an irrotational velocity field $\vec{v}_i = \vec{\nabla} \phi_i$, and a fluid-fluid interface $\xi$ given by $z = z_0 + \xi(x,y,t)$. We then linearise \Eqs{continuity}{euler} around a steady background flow \mbox{ $\vec{v}_0 = \vec{\nabla} \phi_0$} with \mbox{$\phi_i = \phi_0 + \varphi_i$.} We further take $\partial_z B \gg \partial_x B$. At the hard-wall upper and lower boundaries \Eqs{continuity}{BC_kinematic} lead to the Ansatz
\begin{align}
 \varphi_i = \sum_{n} \cosh[\EV_n (z-h_i)] \varphi_{i,n}(x,y,t) \label{eq:Ansatz_phi} ~,
\end{align}
where $n$ labels the eigenfunctions of the 2D-Laplacian $(\nabla^2 + \EV_n^2) \varphi_n = 0$, subject to the boundaries of the interface. The linear equations of motion obtained by \Eql{euler}{BC_dynamic} evaluated at the free interface are
\begin{align}
 \rho_1 \mathcal{D}_{t} \varphi_1 - \rho_2 \mathcal{D}_{t} \varphi_2 &= \left( \sigma \nabla^2 - [\rho] g_0 + \frac{[\chi]}{\mu_0} B \partial_z B \right) \xi \label{eq:EoM_phi}\\
 \mathcal{D}_t \xi &= \frac{1}{2} \partial_z \left( \varphi_1 + \varphi_2 \right) \label{eq:EoM_xi} ~,
\end{align}
where  the curvature for small deformations $\xi$ is given by $(R_1^{-1} + R_2^{-1}) \simeq - \nabla^2 \xi$ \cite{landaufluid} and $\mathcal{D}_t = \partial_t + v_0 \! \cdot \! \nabla$ is the material derivative on the 2D interface.
While the preceding discussion is valid for arbitrary geometries of the interface, flows and heights, we choose for the sake of clarity a plane wave basis (with wavenumber $n=k$ and $\EV_k^2 = k^2$), a vanishing background flow ($\mathcal{D}_t = \partial_t$, depicted with a dot), and equal surface heights $|h_1|=|h_2| \equiv h$.

Since at the interface $v_{1z}=v_{2z}$ (c.f.\ \Eq{BC_kinematic}), we get with $\varphi_k \equiv \varphi_{1,k}$ 
\begin{equation} \label{eq:wave_equation}
 \ddot{\varphi}_k + \omega_k^2 \varphi_k = \frac{\dot{g}_\mathrm{eff}}{G_k} \dot{\varphi}_k ~,
\end{equation}
where
\begin{equation}\label{eq:Gkdef}
G_k = \left( [\rho] g_\mathrm{eff} + \sigma \EVtmp^2 \right) / \tilde{\rho} \,, 
\end{equation}
with $\tilde{\rho} = \rho_1 + \rho_2$.
The left hand side of \Eq{wave_equation} describes the familiar oscillatory behaviour with frequency
\begin{align} \label{eq:full_dispersion}
 \omega_k^2 = G_k \EVtmp \tanh(\EVtmp h)\,.
\end{align}
The effective gravity \cite{BAL15}
\begin{align} \label{eq:g_eff}
  g_\mathrm{eff} =  g - \frac{[\chi]}{[\rho] \mu_0} B \partial_z B  ~,
\end{align}
in our setup can be modulated by the motion of the sample (and thus of the interface $z_0(t)$) through the external magnetic field $B(z)$. 
This introduces an explicit time dependence of the frequency $\omega_k \equiv \omega_k(t)$ as well as an additional friction term to the equation of motion (right hand side of \Eq{wave_equation}).
The acceleration of the sample, $\ddot{z}_0$, can also be taken into account by substituting $g \rightarrow g + \ddot{z}_0$ in \eq{g_eff}.

\textit{Analogue Cosmology.}\textbf{---}In order to make the connection with cosmology transparent we consider the shallow water (or long wavelength) limit $\EVtmp h \ll 1$. The change in the effective gravity \eq{g_eff} directly translates to a change in the propagation speed  $c_k$ of long wavelength perturbations. We define the mode-dependent scale factor 
\begin{align} \label{eq:Gk}
 a_k^{-2}(t) \equiv c_k^2(t) = G_k(t) h ~,
\end{align}
with which the equation of motion \eq{wave_equation} take the form
\begin{align} \label{eq:EoM_Rainbow}
 \ddot{\varphi}_k + 2 \frac{\dot{a}_k}{a_k} \dot{\varphi}_k + \frac{\EVtmp^2}{a_k^2} \varphi_k &= 0 ~.
\end{align}
Thus, our two fluid system in the shallow water limit is equivalent to a massless scalar field in a Friedmann-Lema\^{i}tre-Robertson-Walker (FLRW) type rainbow universe \cite{weinfurtner2009cosmological}. 
In the long wavelength regime, the $k$-dependence of the scale factor 
remains due to 
the surface tension $\sigma$ (c.f.~\Eq{Gkdef}). This regulatory part vanishes when $\sigma \rightarrow 0$, and the effective metric for the perturbations is 
then:
\begin{align} \label{eq:FLRW_metric}
 ds^{2} = g_{\mu\nu} dx^{\mu}dx^{\nu} = -dt^{2} + a^{2}(t) (dx^{2} + dy^{2}) ~.
\end{align}
This is the exact FLRW solution to the Einstein's equations for an expanding, homogeneous and isotropic, universe described by one $k$-independent scale factor $a(t)$.  
In addition, it is possible to change the sign of $G_k(t)$ by inverting the direction of $g_{\rm eff}$ (see Fig. 2A2), previously exploited in magnetic levitation experiments e.g.~\cite{BAL15,berry,geim,liao2017shapes}. 
In our setup, this corresponds to a Hartle-Hawking-like \cite{HartHawk,HartHawkHert} change from Lorentzian to Euclidian signature in the analogue spacetime geometry, as can be seen from Eqs.~(\ref{eq:Gk}) and~(\ref{eq:FLRW_metric}).

Note that even for $\sigma \neq 0$ the ability to change the sign of the effective gravity $g_\mathrm{eff}$ allows for an infinite expansion for a single mode in our system. This can be explicitly seen by calculating the number of e-folds
\begin{align}
 N = \ln\left(\frac{a_k(t_\mathrm{f})}{a_k(t_\mathrm{i})}\right) = \frac{1}{2} \ln\left(  \frac{ \sigma k^{2} +[\rho]g_\mathrm{eff}(t_\mathrm{i}) }{\sigma k^{2} +[\rho] g_\mathrm{eff}(t_\mathrm{f})}  \right) ~,
\end{align}
used in cosmology to describe the length of the inflationary period, where here the denominator can approach zero if $g_\mathrm{eff}$ becomes negative. 
Thus, by changing the magnetic field accordingly, it is possible to engineer the exact time-dependence of the scale factor $a_k(t)$ for the cosmological model one wishes to study. 
This includes, in principle, an arbitrarily high number of e-folds for a fixed value of $k$.

\begin{figure*}[t]	
  \includegraphics[trim=1.7cm 0 0 0]{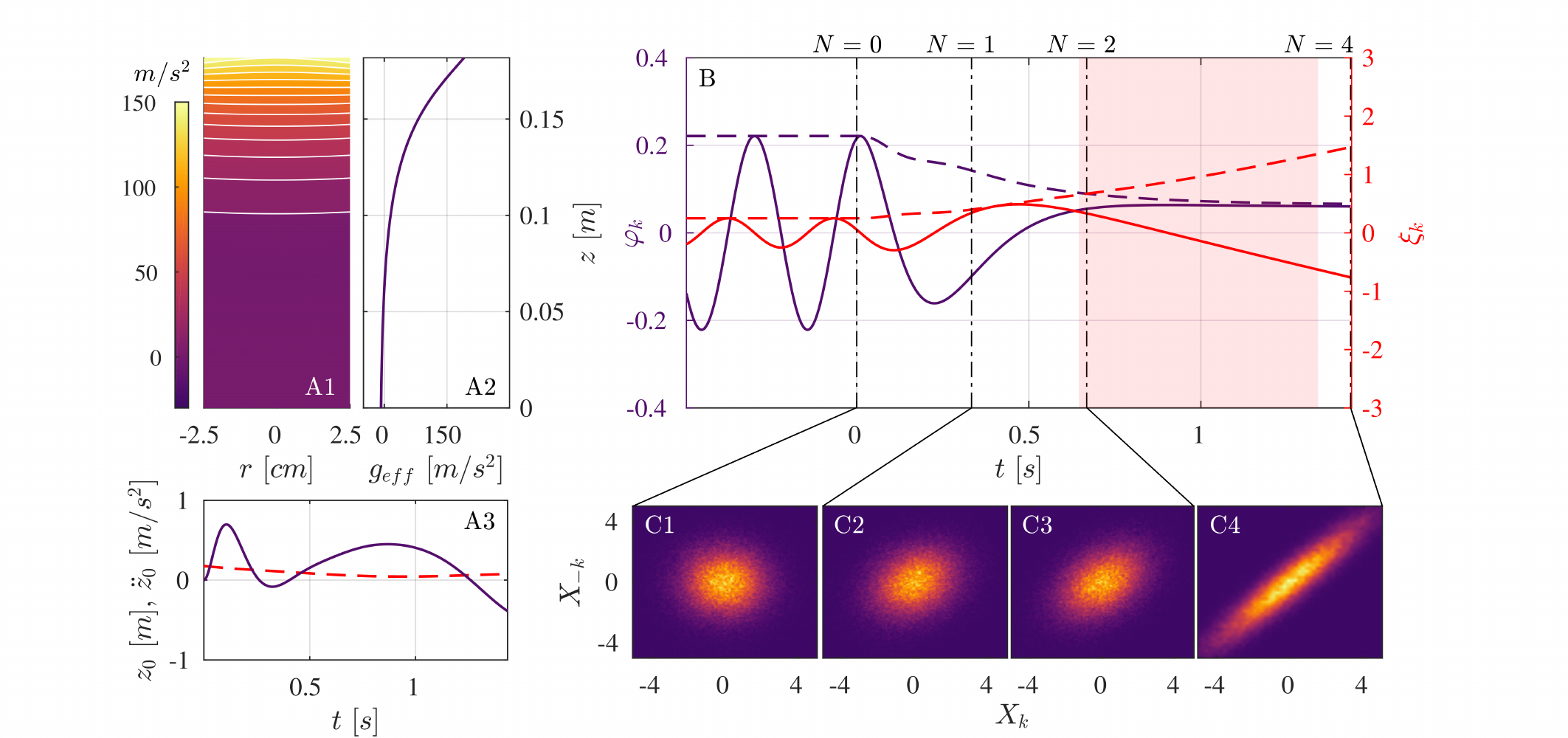}
  \caption{
Panel A1 depicts the effective gravity $g_{\rm eff}$ in the bore of the magnet for the butanol-aqueous solution ($\ddot{z}_0=0$).
The vertical axis gives the vertical position in the magnet $z$, and the horizontal axis the radial position $r$.
The magnitude is given by the colorbar. 
Panel A2 is the effective gravity along the line $r=0$. 
Panel A3 shows the evolution of the position $z_0(t)$ (dashed red) and acceleration $\ddot{z}_0(t)$ (solid purple) of the tank in the magnet in order to have an exponentially inflating effective universe with Hubble parameter $H=3 \, s^{-1}$. 
Panel B depicts the solution to the wave equation \eq{EoM_Rainbow}. 
The solid (dashed) line is the real part (absolute value) of the velocity potential $\phi_k$ (purple) and of the height field $\xi_k$ (red). 
The black dash-dotted lines represent different number of e-folds $N=0,1,2$ and $4$. 
The shaded region indicates where the mode is outside the Hubble horizon. 
Panel C depicts the maximal two-mode squeezing of the system projected onto the instantaneous eigenbasis at the times (equivalently number of e-folds) indicated (see main text for details).}
\label{fig:Fig2}
\end{figure*}

\textit{Experimental implementation.}\textbf{---}In order to minimise the influence of non-linear contributions to the dispersion relation, we choose a two-fluid system consisting of 1-butanol and a weak aqueous paramagnetic solution, with a small interfacial  surface tension, $\sigma=1.8 \mathrm{mN/m}$ \cite{doi:10.1021/ed060pA322.2} (immiscibility requires $\sigma\ne 0$). We take a circular ring-shaped basin for which the azimuthal degrees of freedom obey periodic boundary conditions with a maximal wavelength $\lambda_\mathrm{max} = \pi d$. The diameter $d$ is limited by the size of the bore of the superconducting magnet, which for our experimental setup is $d=4 \mathrm{cm}$. 
A significant advantage of our proposed system is the ability to make non-destructive measurements of interface-waves. For example, using a ``Fast Checkerboard Demodulation" (FCD) method \cite{wildeman2017real} enables the time-resolved measurement of the full dynamics. 
In contrast to the observable universe (which was realised exactly once), in our system we can automise and repeat the experiment a large number of times in order to analyse the statistical properties of the state after inflation. 

We tune the scale factor $a(t)$ to be exponential in the linear dispersion limit. This can be achieved by combining the spatially dependent magnetic forces within the bore with a small mechanical acceleration of the system. In \Fig{Fig2}A1-2 we show the effective gravity in the magnet bore for our butanol-aqueous system. The time-dependent position and acceleration of the basin in the magnet is given in \Fig{Fig2}A3, determined by solving the differential equation $g_\mathrm{eff}[z_0,\ddot{z}_0] = \exp(-2 H t)$, where $H\!\!=\!{\rm const}$ is the effective Hubble parameter.

\textit{Effective horizon and mode freezing.}\textbf{---}One of the hallmarks of inflation, giving rise to the large scale structure formations observed in our universe, is the freezing of modes once they exit the Hubble horizon. It is common to introduce the auxiliary field $\mathcal{X}_k = a_k \, \varphi_k$ for which the wave equation \eq{wave_equation} takes the form of a time-dependent harmonic oscillator with frequency
\begin{align}
 \Omega_k^2(t) = \frac{k^2}{a_k^2} - \frac{\ddot{a}_k}{a_k} ~.
\end{align}
Horizon crossing occurs at $\Omega_k^2 = 0$, separating the oscillating solution dominated by the first term on the right hand side from exponentially growing\,/\,decaying solutions at late times, dominated by the time-independent second term. The essence of inflationary dynamics is fully captured for late times after a mode has crossed the horizon, since the dynamics of the physical field $\varphi_k$ freezes and becomes trivial, obeying a constant solution in time. 

Our system exhibits an analogue behaviour, where each mode crosses the shrinking effective horizon beyond which it freezes. In \Fig{Fig2}B the solution of the field equation \eq{wave_equation} for the longest wavelength of our system is shown for the full non-linear dispersion relation \eq{full_dispersion}. The time evolution is computed numerically using a Runge-Kutta fourth-order scheme. For $t<0$ the system is evolved in flat space, reducing \Eq{wave_equation} to a simple harmonic oscillator. At $t=0$ the system begins to expand, leading to an oscillatory, damped time evolution of the field, which upon crossing the effective horizon approaches rapidly a nearly constant field solution. 
The full model (including dispersive effects) exhibits minor differences to a completely frozen field solution, caused by the surface tension. This introduces a small time-depence to $\ddot{a}_k/a_k$ and in turn a slow further evolution of the field outside the horizon. 
Apart from a different effective expansion experienced by high momentum modes, this regulatory part leads to the mode re-entering the Hubble horizon, exhibiting oscillatory behaviour at later times. 
We further present the evolution of the surface height $\xi_k = a_k^2 \dot{\varphi}_k$, directly accessible in the experiment, which exhibits a growing, non-oscillatory solution after horizon crossing.
The continuous observation of the field dynamics provides direct evidence for the inflationary dynamics of the system.

\textit{Classical two-mode squeezing.}\textbf{---}While the above evolution and mode freezing describe the full dynamics during inflation, there is an equivalent description in terms of squeezed states~\cite{Albrecht:1992kf,Grishchuk:1990cm}. This approach is in close analogy to condensed matter systems undergoing rapid changes (e.g.~quenches~\cite{Hung1213}) in the propagation speed of long-wavelength perturbations. For a quadratic field theory the quantum nature of the system is only determined by the initial conditions, while each trajectory evolves according to the classical equation of motion~\cite{gardiner2004quantum}. Therefore, by emulating the quantum statistics of the initial state, we can study the full time evolution of inflationary physics within the linear regime of fluid dynamics. Our system allows us to stop the expansion and to analyse the inflationary signatures in the resulting flat spacetime through the statistical properties of the state. The associated definition of a well defined ground (vacuum) state leads to cosmological quasi-particle production (mode amplification) and two-mode squeezed states. This is the result of the rapid effective expansion of our analogue universe connecting two flat regions of spacetime $a(t_\mathrm{i}) \to a(t_\mathrm{f})$ \cite{BIR82}. 
Note that while the freezing of the modes is sensitive to $\ddot{a}_k / a_k \approx \mathrm{const}$, these effects are rather generic and occur for more general types of expansion.

In line with quantum field theory in curved spacetime, we introduce the classical quasi-particle amplitudes $b_k$ in the initial flat region of spacetime by expanding the field
$ \mathcal{X}_k(x,t) = \left( f_k^\mathrm{i}(t) b_k + {f_k^\mathrm{i}(t)}^{*} b_{-k}^{*} \right) \exp(\mathrm{i} k x)$,\,
in terms of the time-dependent mode functions $(f_k, f_k^{*})$, normalized by the Wronskian
 \mbox{$\langle f_k \, ; \, f_k \rangle \equiv \mathrm{i} \left( f_k^{*} \partial_t f_k - (\partial_t f_k^{*}) f_k \right) = 1$.} 
As the conservation of the Wronskian implies
\mbox{$|\alpha_k|^2 - |\beta_k|^2 = 1$}, the initial and final flat regions of spacetime are related by a Bogoliubov transformation \mbox{$ f_k^\mathrm{i}(t) = \alpha_k f_k^\mathrm{f}(t) + \beta_k {f_{-k}^\mathrm{f}(t)}^{*}$}. The final state is fully described by the corresponding transformation of the quasi-particle amplitudes \mbox{$d_k \equiv \langle f^\mathrm{f}_k(t) \, ; \, \mathcal{X}_k \rangle = \alpha_k b_k + \beta_k^* b_{-k}^{*}$.} 
The measurable mode intensity after the effective expansion in our system is
\begin{align} \label{eq:amplification}
 \langle d_k^{*} d_k \rangle = \left( 2 |\beta_k|^2 + 1 \right) \langle b_k^{*} b_k \rangle ~,
\end{align}
where $\langle \dots \rangle$ denotes the classical expectation value over sufficiently many measurements, allowing us to assume $\langle b_{-k}^{*} b_{-k} \rangle = \langle b_k^{*} b_k \rangle$. 

\begin{figure}[!t]	
  \includegraphics[width=1\columnwidth]{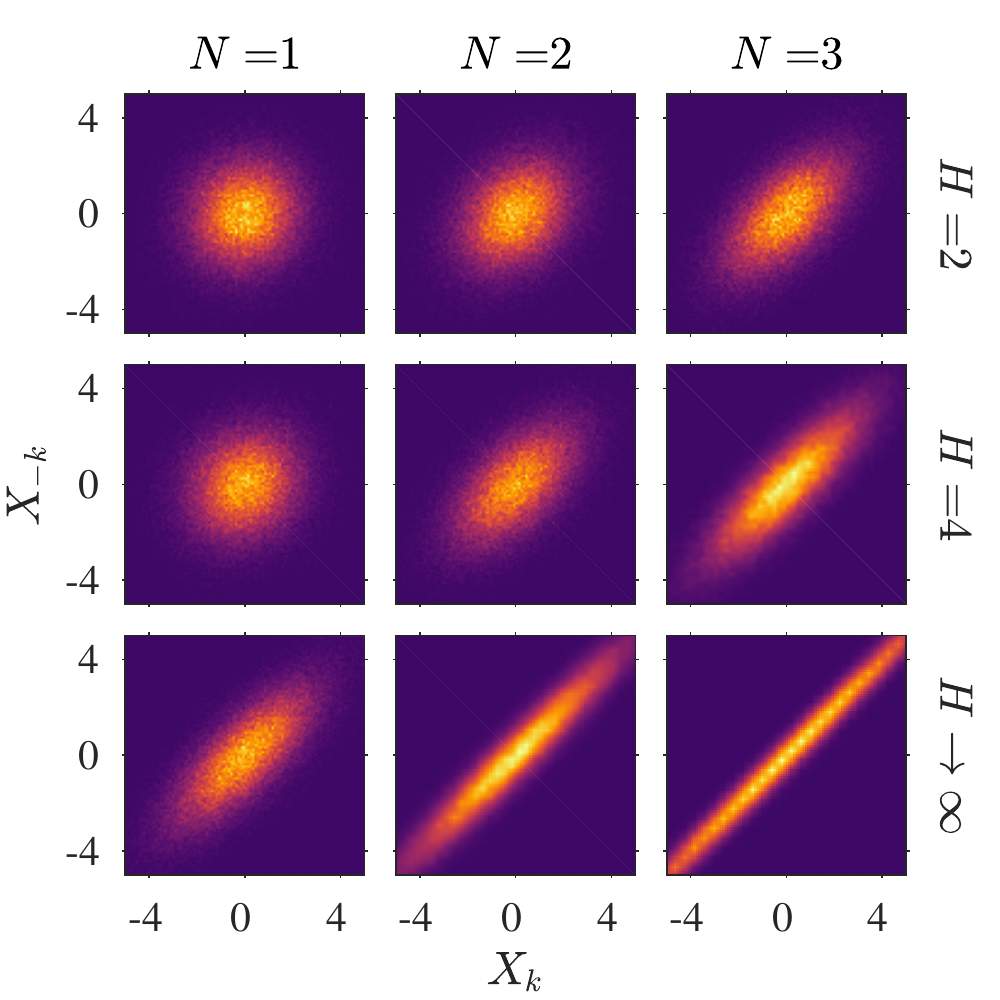}
  \caption{
  Two-mode squeezed states of the field position quadrature  $X_{\pm k}$ (see main text for details).
  Each panel represents the correlation between $X_{k}$ and $X_{-k}$ for different simulation parameters.
  Brighter regions correspond to higher probability of measurement. 
  The number of e-folds increases with each panel from left to right, and the Hubble parameter increases from top to bottom. 
  The final row corresponds to a sudden change (or quench) in the analogue scale factor.
  } \label{FIG3}
\end{figure}

As anticipated from translational invariance in equation~\eqref{eq:EoM_Rainbow}, the Bogoliubov coefficients $\alpha_k$, $\beta_k$ only mix modes of opposite momenta:
\begin{align} \label{eq:off_diag_b_correlators}
 \langle d_{-k} d_k \rangle = 2 \alpha_k \beta_k^{*} \, \langle b_k^{*} b_k \rangle ~.
\end{align}
Quasi-particle amplification (creation) therefore occurs in the form of correlated (entangled), counter-propagating pairs, and our system exhibits the classical analogue of two-mode squeezing. Therein the fluctuations in one quadrature are lowered below their initial value at the cost of increasing the fluctuations in the other. 
Defining the position variable \mbox{$X_{k} = \left( f_{k}^\mathrm{f}(t)d_{k} + f_{k}^\mathrm{f}(t)^*d_{k}^* \right)/|f_{k}^\mathrm{f}|$,} 
we show in \Fig{Fig2}C the two-mode squeezed states for varying number of e-folds $N$ during the expansion. 
For each parameter we sampled $10^5$ trajectories from an initial uncorrelated Gaussian state. 
At the times presented, the maximum squeezing $|\alpha_k - \beta_k|^2$ caused by the correlations~\eq{off_diag_b_correlators} increases with $N$, due to the increasing number of created quasi-particle pairs. A similar behaviour is found by increasing the effective Hubble parameter $H$, as shown in Fig.~\ref{FIG3} for $H=2,4$ $s^{-1}$ and the limit of an instantaneous quench, $H \rightarrow \infty$. 
Note that each mode itself shows no sign of squeezing and is simply amplified according to \Eq{amplification}. 

\textit{Conclusion.}\textbf{---}The precise controllability of the effective metric in our novel system together with the continuous non-destructive measurement of the field dynamics makes it a versatile tool to explore fundamental questions of free (quantum) fields in time-dependent curved spacetimes. 
In particular, the proposed analogue system opens up the possibility to also emulate alternative theories to inflation, such as cyclic and ekpyrotic models~\cite{CyclicSc,Cyclic,Khoury,Lehners,Gasperini,Veneziano}, or even proposals exhibiting a signature change in the spacetime metric~\cite{HawkMoss,Vilenkin1,Vilenkin2,LindeQC,HartHawk,HartHawkHert}. 
Our system has the potential to elucidate the universality and robustness of characteristic cosmological effects. 
For example, it would provide an analogue simulator for evolving field fluctuations through bounces and/or signature changes -- an open problem in cyclic and pre-big bang models. 
This work will advance the mutually beneficial interconnection of cosmological and analogue gravity systems.

\acknowledgements
\emph{Acknowledgements.}
The authors would like to thank 		
J. Berges, M.~V.~Berry, E.~J.~Copeland, M.~Fink, E.~Fort, S.~W.~Hawking, M.~M.~Scase, W.~G.~Unruh and S.~Wildeman for helpful discussions. AA, RH and SW acknowledge the financial support provided by the University of Nottingham through the Research Priority Area Development Fund `Cosmology in a Superconducting Magnet'. AA and SW acknowledge partial support from STFC consolidated grant No. ST/P000703/. SW acknowledges financial support provided under the Royal Society University Research Fellow (UF120112), the Nottingham Advanced Research Fellow (A2RHS2), the Royal Society Project Grant (RG130377), the Royal Society Enhancement Grant (RGF/EA/180286), and the EPSRC Project Grant (EP/P00637X/1). 


	

\begin{thebibliography}{99}

\bibitem{Guth}
A.~H. Guth.
\newblock {The Inflationary Universe: A Possible Solution to the Horizon and
  Flatness Problems}.
\newblock {\em Phys. Rev. D}, 23:347--356, 1981.

\bibitem{LindeNewInf}
A.~D. Linde.
\newblock {A New Inflationary Universe Scenario: A Possible Solution of the
  Horizon, Flatness, Homogeneity, Isotropy and Primordial Monopole Problems}.
\newblock {\em Phys. Lett. }, 108B:389--393, 1982.

\bibitem{MukhChib}
V.~F. Mukhanov and G.~V. Chibisov.
\newblock {Quantum Fluctuations and a Nonsingular Universe}.
\newblock {\em JETP Lett.}, 33:532--535, 1981.
\newblock [Pisma Zh. Eksp. Teor. Fiz.33,549(1981)].

\bibitem{Sasaki}
M.~Sasaki.
\newblock {Gauge Invariant Scalar Perturbations in the New Inflationary
  Universe}.
\newblock {\em Prog. Theor. Phys.}, 70:394, 1983.

\bibitem{GuthPi}
A.~H. Guth and S.~Y. Pi.
\newblock {Fluctuations in the New Inflationary Universe}.
\newblock {\em Phys. Rev. Lett.}, 49:1110--1113, 1982.

\bibitem{Hawking}
S.~W. Hawking.
\newblock {The Development of Irregularities in a Single Bubble Inflationary
  Universe}.
\newblock {\em Phys. Lett.}, 115B:295, 1982.

\bibitem{Starobinsky}
A.~A. Starobinsky.
\newblock {Dynamics of Phase Transition in the New Inflationary Universe
  Scenario and Generation of Perturbations}.
\newblock {\em Phys. Lett.}, 117B:175--178, 1982.

\bibitem{BardSteinTurn}
J.~M. Bardeen, P.~J. Steinhardt, and M.~S. Turner.
\newblock {Spontaneous Creation of Almost Scale - Free Density Perturbations in
  an Inflationary Universe}.
\newblock {\em Phys. Rev. D}, 28:679, 1983.

\bibitem{BAL15}
K.~A. Baldwin, M.~M. Scase, and R.~J.~A. Hill.
\newblock The inhibition of the rayleigh-taylor instability by rotation.
\newblock {\em Scientific reports}, 5:11706, 2015.

\bibitem{UNR81}
W.~G. Unruh.
\newblock {Experimental black hole evaporation}.
\newblock {\em Phys. Rev. Lett.}, 46:1351--1353, 1981.

\bibitem{WEI11}
S.~Weinfurtner, E.~W. Tedford, M.~C.~J. Penrice, W.~G. Unruh, and G.~A.
  Lawrence.
\newblock Measurement of stimulated hawking emission in an analogue system.
\newblock {\em Phys. Rev. Lett.}, 106:021302, Jan 2011.

\bibitem{belgiorno2010hawking}
F.~Belgiorno, S.~L. Cacciatori, M.~Clerici, V.~Gorini, G.~Ortenzi, L.~Rizzi,
  E.~Rubino, V.~G. Sala, and D.~Faccio.
\newblock Hawking radiation from ultrashort laser pulse filaments.
\newblock {\em "Phys. Rev. Lett."}, 105(20):203901, 2010.

\bibitem{euve2016observation}
L.-P. Euv{\'e}, F.~Michel, R.~Parentani, T.~G. Philbin, and G~Rousseaux.
\newblock Observation of noise correlated by the hawking effect in a water
  tank.
\newblock {\em "Phys. Rev. Lett."}, 117(12):121301, 2016.

\bibitem{steinhauer2016observation}
J.~Steinhauer.
\newblock Observation of quantum hawking radiation and its entanglement in an
  analogue black hole.
\newblock {\em Nature Physics}, 12(10):959--965, 2016.

\bibitem{torres2017rotational}
T.~Torres, S.~Patrick, A.~Coutant, M.~Richartz, E.~W. Tedford, and
  S.~Weinfurtner.
\newblock Rotational superradiant scattering in a vortex flow.
\newblock {\em Nature Physics}, 2017.

\bibitem{vocke2017rotating}
D.~Vocke, C.~Maitland, A.~Prain, F.~Biancalana, F.~Marino, and D.~Faccio.
\newblock Rotating black hole geometries in a two-dimensional photon
  superfluid.
\newblock {\em arXiv preprint arXiv:1709.04293}, 2017.

\bibitem{fedichev2004cosmological}
P.~O. Fedichev and U.~R. Fischer.
\newblock “cosmological” quasiparticle production in harmonically trapped
  superfluid gases.
\newblock {\em Phys. Rev. A}, 69(3):033602, 2004.

\bibitem{barcelo2003analogue}
C.~Barcelo, S.~Liberati, and M.~Visser.
\newblock Analogue models for frw cosmologies.
\newblock {\em International Journal of Modern Physics D}, 12(09):1641--1649,
  2003.

\bibitem{weinfurtner2009cosmological}
S.~Weinfurtner, P.~Jain, M.~Visser, and C.~W. Gardiner.
\newblock Cosmological particle production in emergent rainbow spacetimes.
\newblock {\em Classical and Quantum Gravity}, 26(6):065012, 2009.

\bibitem{jain2007analog}
P.~Jain, S.~Weinfurtner, M.~Visser, and C.~W. Gardiner.
\newblock Analog model of a friedmann-robertson-walker universe in
  bose-einstein condensates: Application of the classical field method.
\newblock {\em Phys. Rev. A}, 76(3):033616, 2007.

\bibitem{cha2017probing}
S.-Y. Ch{\"a} and U.~R Fischer.
\newblock Probing the scale invariance of the inflationary power spectrum in
  expanding quasi-two-dimensional dipolar condensates.
\newblock {\em "Phys. Rev. Lett."}, 118(13):130404, 2017.

\bibitem{jaskula2012acoustic}
J.-C. Jaskula, G.~B. Partridge, M.~Bonneau, R.~Lopes, J.~Ruaudel, D.~Boiron,
  and C.~I. Westbrook.
\newblock Acoustic analog to the dynamical casimir effect in a bose-einstein
  condensate.
\newblock {\em "Phys. Rev. Lett."}, 109(22):220401, 2012.

\bibitem{eckel2017supersonically}
S.~Eckel, A.~Kumar, T.~Jacobson, I.~B. Spielman, and G.~K. Campbell.
\newblock A supersonically expanding bose-einstein condensate: an expanding
  universe in the lab.
\newblock {\em arXiv preprint arXiv:1710.05800}, 2017.

\bibitem{berry}
M.~V. Berry and A.~K. Geim.
\newblock Of flying frogs and levitrons.
\newblock {\em European Journal of Physics}, 18(4):307, 1997.

\bibitem{geim}
M.~D. Simon and A.~K. Geim.
\newblock Diamagnetic levitation: flying frogs and floating magnets.
\newblock {\em Journal of Applied Physics}, 87(9):6200--6204, 2000.

\bibitem{RichardArxiv}
M.~M. {Scase}, K.~A. {Baldwin}, and R.~J.~A. {Hill}.
\newblock {The Rotating Rayleigh-Taylor instability}.
\newblock {\em arXiv:1603.00675}.

\bibitem{landaufluid}
L.~D. Landau and E.~M. Lifshits.
\newblock {\em Fluid Mechanics: Transl. from the Russian by JB Sykes and WH
  Reid}.
\newblock Addison-Wesley, 1959.


\bibitem{liao2017shapes}
L.~Liao and R.~J.~A. Hill.
\newblock Shapes and fissility of highly charged and rapidly rotating levitated
  liquid drops.
\newblock {\em "Phys. Rev. Lett."}, 119(11):114501, 2017.

\bibitem{HartHawk}
J.~B. Hartle and S.~W. Hawking.
\newblock {Wave Function of the Universe}.
\newblock {\em Phys. Rev. D}, 28:2960--2975, 1983.

\bibitem{HartHawkHert}
J.~B. Hartle, S.~W. Hawking, and T.~Hertog.
\newblock {No-Boundary Measure of the Universe}.
\newblock {\em Phys. Rev. Lett.}, 100:201301, 2008.

\bibitem{doi:10.1021/ed060pA322.2}
J.~W. Whalen.
\newblock Physical chemistry of surfaces, fourth edition (adamson, arthur w.).
\newblock {\em Journal of Chemical Education}, 60(11):A322, 1983.

\bibitem{wildeman2017real}
S.~Wildeman.
\newblock Real-time quantitative schlieren imaging by fast fourier demodulation
  of a checkered backdrop.
\newblock {\em arXiv preprint arXiv:1712.05679}, 2017.

\bibitem{Albrecht:1992kf}
A.~Albrecht, P.~Ferreira, M.~Joyce, and T.~Prokopec.
\newblock {Inflation and squeezed quantum states}.
\newblock {\em Phys. Rev. D}, 50:4807--4820, 1994.

\bibitem{Grishchuk:1990cm}
L.~P. Grishchuk and Y.~V. Sidorov.
\newblock {Squeezed quantum states in theory of cosmological perturbations}.
\newblock In {\em {5th Seminar on Quantum Gravity Moscow, USSR, May 28-June 1,
  1990}}, pages 678--688, 1990.

\bibitem{Hung1213}
C.-L. Hung, V.~Gurarie, and C.~Chin.
\newblock From cosmology to cold atoms: Observation of sakharov oscillations in
  a quenched atomic superfluid.
\newblock {\em Science}, 341(6151):1213--1215, 2013.

\bibitem{gardiner2004quantum}
C.~Gardiner and P.~Zoller.
\newblock {\em Quantum noise: a handbook of Markovian and non-Markovian quantum
  stochastic methods with applications to quantum optics}, volume~56.
\newblock Springer Science \& Business Media, 2004.

\bibitem{BIR82}
N.~D. Birrell and P.~C.~W. Davies.
\newblock {\em Quantum Fields in Curved Space}.
\newblock Cambridge Monographs on Mathematical Physics. Cambridge University
  Press, 1982.

\bibitem{CyclicSc}
P.~J. Steinhardt and N.~Turok.
\newblock {A Cyclic model of the universe}.
\newblock {\em Science}, 296:1436--1439, 2002.

\bibitem{Cyclic}
P.~J. Steinhardt and N.~Turok.
\newblock {Cosmic evolution in a cyclic universe}.
\newblock {\em Phys. Rev. D}, 65:126003, 2002.

\bibitem{Khoury}
J.~Khoury, B.~A. Ovrut, P.~J. Steinhardt, and N.~Turok.
\newblock {The Ekpyrotic universe: Colliding branes and the origin of the hot
  big bang}.
\newblock {\em Phys. Rev. D}, 64:123522, 2001.

\bibitem{Lehners}
J.-L. Lehners.
\newblock {Ekpyrotic and Cyclic Cosmology}.
\newblock {\em Phys. Rept.}, 465:223--263, 2008.

\bibitem{Gasperini}
M.~Gasperini and G.~Veneziano.
\newblock {The Pre - big bang scenario in string cosmology}.
\newblock {\em Phys. Rept.}, 373:1--212, 2003.

\bibitem{Veneziano}
G.~Veneziano.
\newblock {Scale factor duality for classical and quantum strings}.
\newblock {\em Phys. Lett. }, 265B:287--294, 1991.

\bibitem{HawkMoss}
S.~W. Hawking and I.~G. Moss.
\newblock {Supercooled Phase Transitions in the Very Early Universe}.
\newblock {\em Phys. Lett.}, 110B:35--38, 1982.

\bibitem{Vilenkin1}
A.~Vilenkin.
\newblock {Creation of Universes from Nothing}.
\newblock {\em Phys. Lett.}, 117B:25--28, 1982.

\bibitem{Vilenkin2}
A.~Vilenkin.
\newblock {The Birth of Inflationary Universes}.
\newblock {\em Phys. Rev. D}, 27:2848, 1983.

\bibitem{LindeQC}
A.~D. Linde.
\newblock {Quantum Creation of the Inflationary Universe}.
\newblock {\em Lett. Nuovo Cim.}, 39:401--405, 1984.
\end{thebibliography}
\end{document}